\documentclass[journal,twoside,web]{ieeecolor}
\usepackage{tmi}
\usepackage{booktabs}
\usepackage{cite}
\usepackage{amsmath,amssymb,amsfonts}
\usepackage{algorithmic}
\usepackage{dblfloatfix}
\usepackage{graphicx}
\usepackage{hyperref}
\usepackage{textcomp}
\usepackage{caption}
\usepackage{subcaption}
\usepackage{atbegshi}
\usepackage{xstring}
\usepackage{flushend}
\usepackage[normalem]{ulem}
\usepackage{tikz}

\definecolor{darkred}{rgb}{.8,.0,.0}
\newcommand\copyrighttext{%
  \footnotesize \textcopyright 2023 IEEE. Personal use of this material is permitted. Permission from IEEE must be obtained for all other uses, including republication/redistribution.}
\newcommand\copyrightnotice{%
\begin{tikzpicture}[remember picture,overlay]
\node[anchor=south,yshift=10pt] at (current page.south) {\fbox{\parbox{\dimexpr\textwidth-\fboxsep-\fboxrule\relax}{\copyrighttext}}};
\end{tikzpicture}%
}

\AtBeginDocument{\AtBeginShipoutNext{\AtBeginShipoutDiscard}}
\hypersetup{
    colorlinks=true,
    linkcolor=blue,
    filecolor=magenta,      
    urlcolor=cyan,
    pdftitle={Coronary Plaque Meshes},
    }
    
\def\BibTeX{{\rm B\kern-.05em{\sc i\kern-.025em b}\kern-.08em
    T\kern-.1667em\lower.7ex\hbox{E}\kern-.125emX}}
\begin{document}
\title{Automatic Coronary Artery Plaque Quantification and CAD-RADS Prediction using Mesh Priors}

\author{Rudolf L. M. van Herten, Nils Hampe, Richard A. P. Takx, Klaas Jan Franssen, Yining Wang, \\ Dominika Such\'a, Jos\'e P. Henriques, Tim Leiner, R. Nils Planken, Ivana I\v{s}gum\\}
\thanks{Manuscript received December 23rd, 2022; revised May 19th 2023; revised August 29th 2023; accepted October 13th, 2023. This study was supported by the DLMedIA program (P15-26) funded by Dutch Technology Foundation with participation of Philips Healthcare. This study was also partially funded by the DEBuT-LRP TKI-PPP grant (Grant No. NCT04765956) \textit{(Corresponding author: Rudolf L.M. van Herten)}.}
\thanks{R.L.M. van Herten, N. Hampe, and I. I\v{s}gum are with the Department of Biomedical Engineering and Physics, Amsterdam Cardiovascular Sciences, Amsterdam UMC, 1105 AZ Amsterdam, The Netherlands, and also with the Informatics Insitute, University of Amsterdam, 1090GH Amsterdam, The Netherlands (e-mail: r.l.m.vanherten@amsterdamumc.nl; n.hampe@amsterdamumc.nl; i.isgum@amsterdamumc.nl).}
\thanks{R.A.P. Takx, K.J. Franssen, R.N. Planken, and I. I\v{s}gum are with the Department of Radiology and Nuclear Medicine, Amsterdam UMC, 1105 AZ Amsterdam, The Netherlands (e-mail: r.takx@amsterdamumc.nl; k.j.franssen@amsterdamumc.nl; r.n.planken@amsterdamumc.nl).}
\thanks{D. Such\'a and T. Leiner are with the Department of Radiology, UMC Utrecht, University of Utrecht, 3508 GA Utrecht, The Netherlands. T. Leiner is also with the Department of Radiology, Mayo Clinic, Rochester, MN, United States (e-mail: d.sucha-3@umcutrecht.nl; tleiner2@umcutrecht.nl).}
\thanks{J.P. Henriques is with the Heart Centre, Academic Medical Centre, Amsterdam Cardiovascular Sciences, Heart failure \& arrhythmias, 1105 AZ Amsterdam, The Netherlands (e-mail: j.p.henriques@amsterdamumc.nl)}
\thanks{Y. Wang is with the Department of Radiology, Peking Union Medical College Hospital, Chinese Academy of Medical Sciences and Peking Union Medical College, 100730 Beijing, China (e-mail: wangyining@pumch.cn).}

\maketitle
\copyrightnotice
\begin{abstract}
Coronary artery disease (CAD) remains the leading cause of death worldwide. Patients with suspected CAD undergo coronary CT angiography (CCTA) to evaluate the risk of cardiovascular events and determine the treatment. Clinical analysis of coronary arteries in CCTA comprises the identification of atherosclerotic plaque, as well as the grading of any coronary artery stenosis typically obtained through the CAD-Reporting and Data System (CAD-RADS). This requires analysis of the coronary lumen and plaque. While voxel-wise segmentation is a commonly used approach in various segmentation tasks, it does not guarantee topologically plausible shapes. To address this, in this work, we propose to directly infer surface meshes for coronary artery lumen and plaque based on a centerline prior and use it in the downstream task of CAD-RADS scoring. The method is developed and evaluated using a total of 2407 CCTA scans. Our method achieved lesion-wise volume intraclass correlation coefficients of 0.98, 0.79, and 0.85 for calcified, non-calcified, and total plaque volume respectively. Patient-level CAD-RADS categorization was evaluated on a representative hold-out test set of 300 scans, for which the achieved linearly weighted kappa ($\kappa$) was 0.75. CAD-RADS categorization on the set of 658 scans from another hospital and scanner led to a $\kappa$ of 0.71. The results demonstrate that direct inference of coronary artery meshes for lumen and plaque is feasible, and allows for the automated prediction of routinely performed CAD-RADS categorization.
\end{abstract}

\begin{IEEEkeywords}
Convolutional neural network, coronary artery plaque, coronary CT angiography, CAD-RADS, mesh generation
\end{IEEEkeywords}

\section{Introduction}
\label{sec:introduction}
\IEEEPARstart{C}{oronary} CT angiography (CCTA) is an established technique for the non-invasive identification and exclusion of patients with suspected coronary artery disease (CAD)\cite{Hoffmann2012, Kirisli2013}, which remains the leading cause of death worldwide\cite{Mozaffarian2016, WHO}. Typically, clinical assessment is performed through visual or semi-automatic analysis of the CCTA image \cite{Cury2016a, maroules2018coronary}. In visual inspection an expert grades the severity of any stenosis, i.e. the degree of coronary artery lumen narrowing, and identifies atherosclerotic plaque composition, location, and size. Atherosclerotic plaque is categorized into calcified plaque (CP), non-calcified plaque (NCP), and mixed plaque, which is a mixture of the two plaque types \cite{Raff2009}. Contrast enhancement enables good visualization of the coronary artery lumen and NCP while allowing for CP quantification despite partially overlapping density between the contrast medium and calcifications \cite{wolterink2015automatic, Wolterink2016a}. The build-up of atherosclerotic plaque may lead to stenosis. A high stenosis degree may limit blood supply to the myocardium, potentially causing myocardial ischemia \cite{Falk1995}. Plaque types further differ in stability, meaning that it is crucial to distinguish coronary artery plaque types and stenoses. The CAD-Reporting and Data System (CAD-RADS) is used to guide the management of patients with suspected CAD. This is a structured system that grades the disease severity on a 0 to 5 scale (0\%, 1-24\%, 25-49\%, 50-69\%, 70-99\%, and 100\% stenosis). The patient-level score is determined according to the most severe stenosis found among all coronary arteries in the CCTA scan, and may include modifiers depending on stenosis locations and image quality \cite{Cury2016a}.

Early methods analyzing coronary artery plaque and stenosis focused on the (semi-automatic) classification and quantification of coronary artery plaque, which are typically cumbersome and time-consuming tasks requiring a high level of expertise, suffering from high inter-observer variability \cite{jonas2022interobserver}. To improve plaque segmentation, density-level thresholding using the lumen intensity in the aorta was initially examined but showed considerable variability, especially for the detection of NCP \cite{kristanto2013meta, Schepis2010, Dey2010}. More sophisticated methods were therefore proposed to segment NCP. These typically evaluate the multi-planar reformatted (MPR) images derived from CCTA volumes, which allow better artery visualization. Several authors proposed to leverage image derivatives in combination with conventional image processing to create lumen and vessel wall contours \cite{boogers2012automated, ghanem2019automatic}. However, both thresholding and conventional methods failed to produce adequate quantification without requiring substantial expert interaction. More recently, the quantification of NCP has been investigated with deep learning methods. Lin \textit{et al.} \cite{Lin2022} used a hierarchical recurrent convolutional neural network (RCNN), which segmented the outer vessel wall and vessel lumen and CP in two separate optimization schemes. Predictions were then merged to produce lumen, CP, and NCP segmentations.

Unlike segmentation of NCP, segmentation of CP is less complex. Several methods have segmented CP by analyzing CCTA volumes through deep learning approaches, and reached performance approaching interobserver agreement \cite{Wolterink2016, fischer2020accuracy, zhai2022learning}. 

\begin{figure}[t]
\centering
     \includegraphics[trim=0cm 0.1cm 0cm 0cm, width=0.49\textwidth]{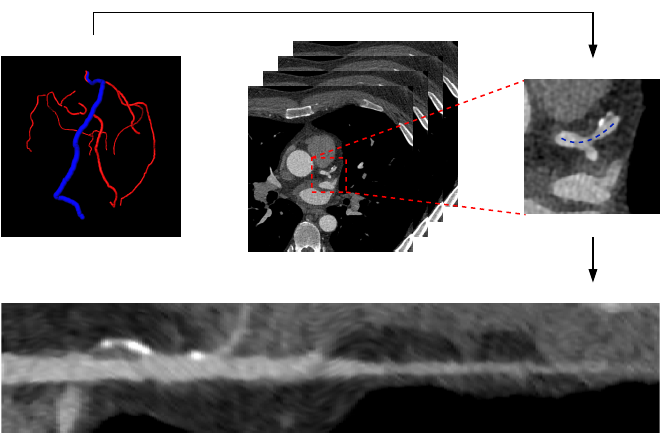}
      \caption{Coronary artery centerline extraction pipeline. Centerlines are extracted from the CCTA image through a previously described automatic method \cite{Wolterink2019}. This example presents how the left anterior descending artery (blue) is subsequently selected from the detected centerlines, from which a straightened MPR volume is reconstructed (bottom) \cite{gupta2020performance}.}
       \label{data}
\end{figure}

Several methods have focused on automatic stenosis detection and CAD-RADS scoring. One of the first approaches was proposed by Boogers \textit{et al.}, who used a minimum cost approach to identify the lumen contour and calculate stenosis by quantifying deviation from the reference contour \cite{boogers2010automated}. A machine learning method was proposed by Zuluaga \textit{et al.} \cite{Zuluaga2011}, who considered MPR slices with lesions to be local outliers compared to slices featuring healthy coronary artery segments. A support vector machine was subsequently used to detect outliers based on features derived from MPRs. Denzinger \textit{et al.} proposed to use a 2D convolutional neural network (CNN) that takes a longitudinal slice of an MPR volume as input to compute a hidden representation \cite{Denzinger2020, Denzinger2022}. This process is repeated for different angles around the coronary artery centerline, after which a small neural network aggregates the outputs from the different angles and predicts the final stenosis degree.

Since plaque and stenosis assessment are highly correlated, several methods combined plaque categorization or quantification with stenosis detection. These methods frequently featured recurrent convolutional neural networks (RCNNs), allowing them to capture spatial information by learning the dependencies of sequential MPR slices. Zreik \textit{et al.} \cite{Zreik2018} employed an RCNN operating on MPR volumes to both detect plaque and determine lesion-wise stenosis severity. Similarly, Lin \textit{et al.} used their segmentation RCNN in the downstream task of stenosis grading by performing rule-based CAD-RADS grading \cite{Lin2022}.

Recently, coronary lumen surface meshes have been investigated as the precursor for stenosis detection and grading. Such meshes present the lumen as a contiguous structure with sub-voxel accuracy, which is not provided by voxel-based segmentation. Therefore, methods have been developed to directly infer a surface mesh based on deforming a shape prior with image features.
For example, Lugauer \textit{et al.}\cite{lugauer2014precise} adopted a Markov random field formulation for tubular segmentation by providing convex shape priors, thus regularizing lumen segmentation. Sivalingam \textit{et al.}\cite{sivalingam2016inner} encoded contextual image features with a combination of active contour models and random forest regression for the delineation of both the lumen and outer wall.
Lee \textit{et al.}\cite{Lee2019} assumed coronary artery segments to have a roughly tubular shape, which is then deformed to match the visible lumen in the CCTA image using a CNN. More recently, Wolterink \textit{et al.} \cite{Wolterink2019a} proposed graph convolutional networks (GCNs) to predict the exact location of vertices along the coronary artery luminal mesh. This network uses localized image features as well as features of neighboring vertices to produce a smooth tubular surface matching the coronary lumen. In a similar fashion, Alblas et al. \cite{alblas2022deep} proposed to predict lumen and outer wall radii of the carotid arteries in black-blood MRI by accumulating image features with a CNN, outperforming voxel-based alternatives.

To address the issues posed by voxel-based plaque segmentation presented in previous work, we build on recent works performing mesh generation. We introduce a plaque quantification method that leverages the inherent tubular structure of coronary arteries, which allows us to directly infer a mesh for coronary artery lumen and plaque based on a centerline prior. The location of vertices on the surface meshes is directly optimized by regressing the extent of lumen, CP, and NCP present in an intensity profile originating from the coronary artery centerline. Therefore, the method does not suffer from contiguity errors that may cause anatomically implausible shapes and minimizes post-processing for downstream tasks. The result allows direct quantification of CP and NCP. Thereafter, we propose to leverage the obtained meshes to predict the patient-level CAD-RADS grade. As such, our contributions are as follows:
\begin{itemize}
    \item We propose a CNN that leverages anatomical coronary artery centerline priors to create boundaries for the arterial lumen and plaque, resulting in meshes that are directly applicable in downstream tasks.
    \item We provide a deep learning-based method to directly infer patient-level CAD-RADS from the automatically obtained meshes.
    \item We are the first to design a fully automated method that covers the complete clinical coronary artery analysis workflow from lumen and plaque evaluation and quantification to stenosis grading, without the need for voxel-wise segmentation or rule-based post-processing.
\end{itemize}

The remainder of this manuscript is structured as follows: Section~\ref{DataSec} describes the data and reference standard. Section~\ref{MethodSec} describes the methodology, for which the experimental results are covered in Section~\ref{Experiments}. A comparison with previous methods is provided in Section~\ref{CompareSec}, and results are discussed in Section~\ref{DiscSec}.

\section{Data}
\label{DataSec}

\subsection{Patient and Image Data}
This study comprises retrospectively collected CCTA scans obtained from clinical routine of three hospitals. Set 1 originates from the Amsterdam University Medical Center - Location UvA, The Netherlands. Set 2 was acquired in the Peking Union Medical College Hospital, China, and Set 3 was obtained from the University Medical Center Utrecht, The Netherlands. For all sets, the need for informed consent was waived by the (local) Institutional Ethical Review Board. All scans were acquired according to guidelines for coronary CTA as presented by the Society of Cardiovascular Computed Tomography \cite{abbara2016scct}. Furthermore, all patients were administered contrast agent before acquisition, and scans were electrocardiogram-triggered.

Set 1 contains CCTA exams of 1,660 patients acquired on a Siemens Somaton Force CT scanner (Siemens Healthineers, Erlangen, Germany). Tube voltage ranged from 70 to 120 kVp and the tube current ranged from 296 to 644 mAs. Scans had an in-plane resolution of 0.26-0.46 mm\textsuperscript{2} and a slice thickness and increment of 0.6 mm.

For the 46 patients in Set 2, CCTA exams were acquired on an IQon Spectral CT scanner (Philips Healthcare, Best, The Netherlands) with a tube voltage of 120 or 140 kVp and a tube current ranging from 50 to 960 mAs. Additionally, scans had an in-plane resolution of 0.43-0.59 mm\textsuperscript{2} and a slice thickness and increment of 0.8 or 0.9 mm.

Set 3 features 701 CCTA exams of patients acquired on a Brilliance iCT scanner (Philips Healthcare, Best, The Netherlands) with a tube voltage of 80 to 140 kVp and a tube current ranging from 30 to 700 mAs. In-plane resolution was 0.26-0.49 mm\textsuperscript{2} and slice thickness and increment were 0.9 mm.

In addition, we include a final set of 18 public CCTA exams from the Coronary Artery Stenoses Detection and Quantification Evaluation Framework \cite{Kirisli2013} for evaluation and direct comparison with previous work (Set 4).

\begin{figure*}[t]
\centering
     \includegraphics[trim=0.3cm 0cm 0cm 0cm, width=1.\textwidth]{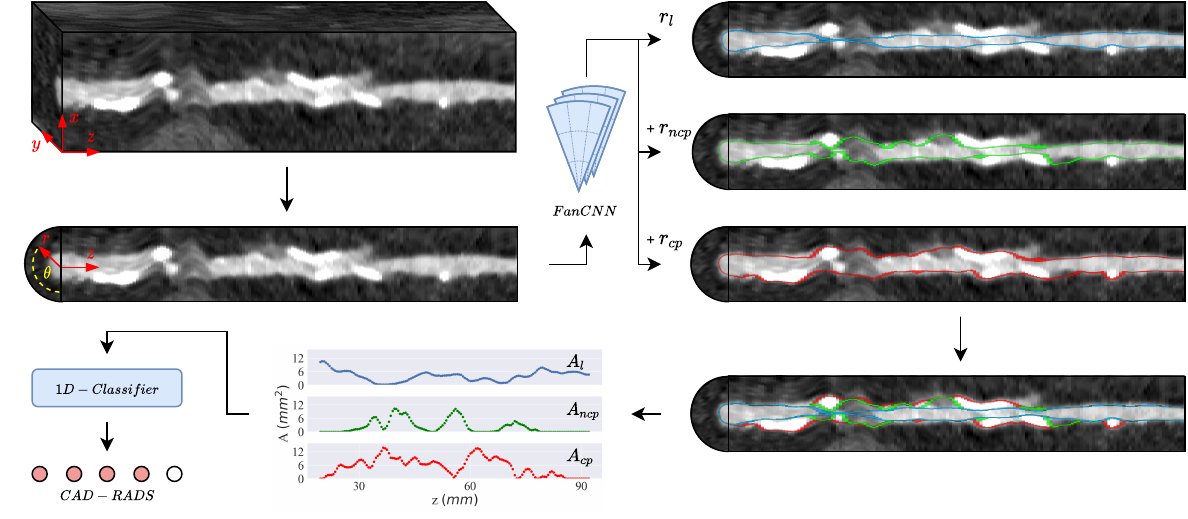}
      \caption{Overview of the proposed methodology. The input MPR volume is first transformed from a Euclidian to a cylindrical coordinate system with the coronary artery centerline defining the longitudinal  ($z$) axis. For a number of predefined equally spaced angles $\theta$, the radial extent of the lumen ($r^v_l$), calcified plaque ($r^v_{cp}$), and non-calcified plaque ($r^v_{ncp}$) is then inferred through a 3D CNN that has learned to operate on cylindrical data (FanCNN). Stacking these radii hierarchically results in a mesh for the coronary lumen ($r_l$), lumen + NCP ($r_l + r_{ncp}$), and lumen + total plaque ($r_l + r_{cp} + r_{ncp}$), the latter of which also defines the outer vessel wall. Cross-sectional area measures are inferred directly from these meshes for each class, and are processed by a 1D Classifier that determines the artery stenosis degree following CAD-RADS categorization (0\%, 1-24\%, 25-49\%, 50-69\%, 70-99\%, 100\%).}
       \label{method}
\end{figure*}

Since the proposed method requires a centerline prior, coronary artery centerlines were extracted for each CCTA using our previously described automatic method exploiting a CNN \cite{Wolterink2019}. Thereafter, the extracted arteries were automatically assigned anatomical labels, resulting in a total of three major coronary centerlines per CCTA, along with any side branches \cite{hampe2021graph}. Centerlines were subsequently used to reconstruct a straightened MPR volume for each artery, with an in-plane resolution of 0.1 mm and a through-plane resolution of 0.5 mm. This resulted in MPR volumes of 12.7 mm $\times$ 12.7 mm $\times$ $0.5L$ mm, with $L$ denoting the variable number of points along the coronary artery centerline. Further analysis was conducted using the obtained MPR volumes. Fig.~\ref{data} provides an overview of the extraction of coronary arteries.

\subsection{Reference Standard}
\subsubsection{Artery and Plaque Segmentation}
Since voxel-wise segmentation of coronary artery plaque to define a reference standard is time-consuming and challenging, only a limited set of CCTA scans was manually annotated. The segmentations provide training data and data for quantitative evaluation of plaque measurements. To ensure diversity in this set, scans were selected based on visual assessment of the amount and type of plaque present in the coronary arteries. Coronary centerlines were further inspected and manually corrected when deemed necessary to create a high quality plaque reference standard. The annotator manually labeled voxels throughout a cross-sectional and two longitudinal planes of the MPR images by voxel-wise labeling, which produced reference labels for lumen, CP, and NCP. Intensity level thresholds were not used, as they typically do not allow for reproducible (NCP) quantification \cite{blackmon2009reproducibility, van2022generative}. To provide training data, a trained observer annotated lumen, CP, and NCP in 59 CCTA exams from Set 1 and 36 from Set 2, totaling 95 CCTA exams. To provide test data and to enable estimation of the interobserver agreement, two experts (Observer 1, Observer 2) annotated a total of 10 coronary arteries from Set 1 (3 RCA, 4 LAD, 3 LCx).

\subsubsection{CAD-RADS Categorization}
The remaining scans from Set 1 and Set 2 and all scans from Set 3 were used to develop and evaluate CAD-RADS categorization. Exclusion criteria for CAD-RADS scoring were a missing or indeterminable CAD-RADS score, patients with stents, and failure of the centerline extraction algorithm. CCTA exams from Set 1 were assigned a patient-level CAD-RADS score as part of standard radiology reporting, while CCTA exams from Set 2 and Set 3 were retrospectively assigned a CAD-RADS score by Observer 1. The grading was performed per stenosis (0 to 5: 0\%, 1-24\%, 25-49\%, 50-69\%, 70-99\%, 100\%), after which per patient CAD-RADS grade was determined according to the most severe stenosis found in all patient arteries. This resulted in dataset sizes of 1,400 for Set 1, 10 for Set 2, and 658 patients for Set 3. From these, 1110 scans from Set 1 were used for training. The remaining 290 scans from Set 1 and the 10 scans from Set 2 were used for testing (hold-out test set), as well as all 658 scans from Set 3 (external test set). Note that scans from Set 2 and Set 3 were only used to evaluate the CAD-RADS categorization algorithm. For all training set scans, it was ensured that the coronary artery centerline containing the most severe stenosis was present in the data. For all test sets, all automatically extracted coronary artery centerlines were used to evaluate CAD-RADS scoring without manual intervention.

\section{Method}
\label{MethodSec}
We present a method that infers a topologically plausible surface mesh for the coronary artery lumen, CP, and NCP. Locations of vertices on the mesh are optimized with a 3D CNN that operates on cylindrical data (\textit{FanCNN}). Cross-sectional area of coronary lumen and plaque is calculated directly from the mesh outputs rendered by this network. These are subsequently processed as 1D signals along the coronary artery centerline by another CNN that predicts stenosis degree according to CAD-RADS categorization. This is performed for all three main coronary arteries, with the highest prediction indicating the patient-level CAD-RADS score. The full overview of this method is presented in Fig.~\ref{method}.

\subsection{Coronary Artery Mesh Generation}
\label{meshSec}
To find the boundaries of the coronary artery lumen and plaque, knowledge about the roughly tubular shape of coronary arteries is exploited. Hence, to incorporate this shape prior, Euclidean topology of the MPR volume is transformed into a cylindrical representation. For this, in each cross-sectional slice $z$ the origin is defined by a point on the coronary artery centerline, while the parameters $r$ and $\theta$ describe the distance and angle at which a point in space is located from the origin respectively. Utilizing this representation, a total of $N_{\theta}$ rays are cast from each pole at equiangular intervals $\theta^v$, resulting in a total of $N_{\theta} \times L$ intensity profiles describing the MPR volume, where $L$ denotes the number of cross-sectional slices. Each intensity profile is effectively a vector of Hounsfield units of size $R$ = 32 and a resolution of 0.2 mm, allowing for the analysis of both the coronary artery and its vicinity \cite{dodge1992lumen}. The intensity profile along with its local neighborhood is the input to the CNN. To allow a CNN to operate on this cylindrical data, the MPR is unfolded into a new 3D volume of size $N_{\theta} \times R \times L$. 

\begin{figure*}[t]
\centering
     \includegraphics[trim=0cm 0.3cm 0cm 0cm, height=4.6cm]{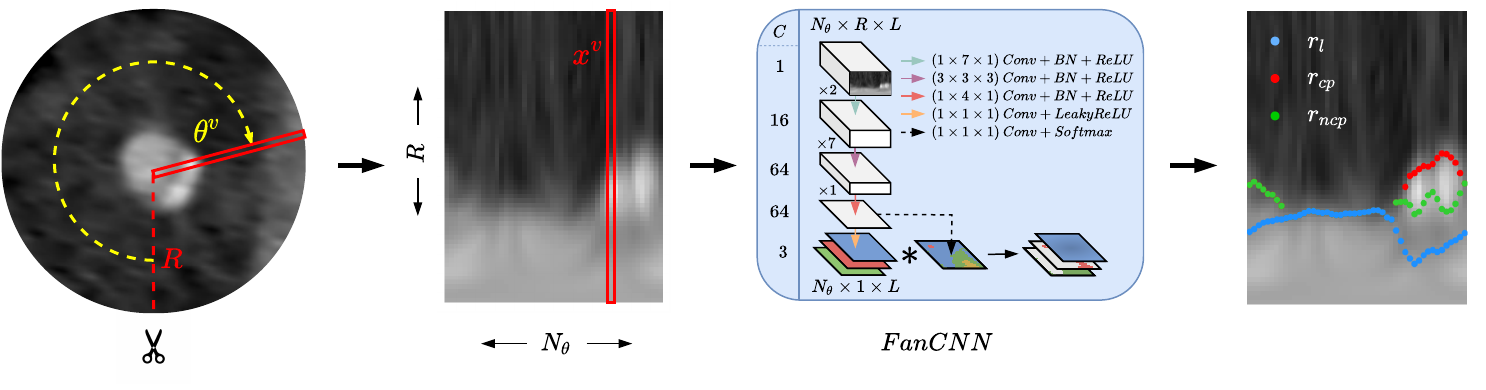}
      \caption{An example depicting the unwrapping and processing of intensity profiles of a cross-section at depth $z$ of an MPR volume. Each intensity profile $x^v$ covers the full radial direction $R$ and is described by an angle $\theta^v$ and depth along the coronary artery centerline $z^v$. By processing each intensity profile in conjunction with its neighborhood in $\theta$ and $z$ direction using the proposed FanCNN, we produce $r_l$, $r_{cp}$, and $r_{ncp}$, which serve as vertices for the lumen and plaque surface meshes. Finally, the plaque outputs are multiplied by a classification head determining the presence and type of plaque (i.e. no plaque, CP, NCP, or mixed plaque), thus ensuring sparsity in the plaque surface mesh.}
       \label{intensity profile}
\end{figure*}

Each intensity profile $x^v$ features a corresponding radial extent for lumen ($r^v_l$), CP ($r^v_{cp}$), and NCP ($r^v_{ncp}$). These are used to generate vertices defining the mesh, and describe how much a vertex needs to be displaced from the coronary artery centerline to reach the lumen and plaque boundaries (see Fig.~\ref{method}). Vertices are thus defined by the cylindrical coordinates ($\theta^v, r^v, z^v$) within the MPR volume, for which $r^v$ is the only unknown. As such, a CNN is employed to reduce the $N_{\theta} \times R \times L$ input to a size of $N_{\theta} \times 3 \times L$, where each output in the radial dimension corresponds to a prediction of one of the radii outputs, optimized through regression. Combining the predictions of all the estimates for all vertices of a single coronary artery results in two meshes describing the coronary artery lumen ($r_l$) and coronary artery outer wall ($r_l + r_{cp} + r_{ncp}$), of which the latter builds on the lumen predictions and can further be decomposed into CP and NCP. A schematic overview of the radius estimation is shown in Fig.~\ref{intensity profile}.

The proposed FanCNN processes the 3D cylindrical data in the described format through a series of 1D and 3D convolution operations. In order to reduce size and ensure a rich set of features in the $R$ direction, two 1D convolutions of kernel size 7 are first employed which generate a total of 16 feature maps per intensity profile. A series of seven 3D convolutions is then used to increase the receptive field for each intensity profile in $\theta$ and $z$ direction, allowing for a better estimate of the radii outputs. This sequence uses a kernel size of $3 \times 3 \times 3$, further reducing the spatial size in the $R$ direction. Circular padding is performed in the $\theta$ direction, while mirror padding is performed in $z$ direction, thus preserving input size in these dimensions. The pre-final layer consists of another 1D convolution of size 4, reducing the size of the $R$ direction to 1. The final layer consists of two output heads: the first head results in three output channels through a linear layer implemented as a $1 \times 1 \times 1$ convolutional kernel, which represents the radii estimates. The second output head performs plaque type classification through a linear layer followed by a softmax activation function. All convolutions are followed by a batch normalization layer and a ReLU activation function, except for the final layer which only employs a LeakyReLU activation.

During inference we employ \textit{classifier attention}, in which predicted lesions are removed if the classification output does not identify any plaque. Classifier attention implicitly classifies all remaining lesions by assigning the same plaque type classification output to all adjacent radii predictions with a radius of at least 0.15 mm.

The loss function employed for this method considers both the classification and regression of each intensity profile $x^v$ and is a summation of several terms constituting different targets:

\begin{equation}
   \mathcal{L} = \mathcal{L}_{cls} + \mathcal{L}_{l} + \mathcal{L}_{cp} + \mathcal{L}_{ncp}
   \label{total loss}
\end{equation}

Here the loss for all samples in a batch is defined, with $\mathcal{L}_{cls}$ being the categorical cross-entropy for the four plaque categories defined above. The loss for the lumen radius $\mathcal{L}_{l}$ is the mean squared error loss:

\begin{equation}
    \mathcal{L}_{cls} = - \frac{1}{K}\sum_{v}\sum_{i}y_i^v\log (p_i^v)
    \label{classifier loss}
\end{equation}
\begin{equation}
    \mathcal{L}_{l} = \frac{1}{K}\sum_{v} (r_l^v -\hat{r}_l^v)^2
    \label{lumen loss}
\end{equation}

where in Eq.~\ref{classifier loss}, $y_i^v$ is a one-hot vector encoding of the plaque labels with $i = (0, 1, 2, 3)$ for lumen, CP, NCP, and the presence of both plaque types (mixed) respectively, while $p_i^v$ is the softmax output probability for the plaque label predicted by the model. We further define the set of all vertices in a batch $K = \{v\}$. For the final two constituent loss terms $\mathcal{L}_{cp}$ and $\mathcal{L}_{ncp}$, the regression loss is calculated as a weighted average of the regions for which the corresponding plaque type is present (see Eq.~\ref{CP loss}, \ref{NCP loss}). For this purpose, two subsets of $K$ are defined for CP and NCP as $C = \{v: r_{cp}^v >0\}$ and $S = \{v: r_{ncp}^v >0\}$ respectively.

\begin{align}
\mathcal{L}_{cp} = & \frac{1}{C}\sum_{v \in C} (r_{cp}^v - \hat{r}_{cp}^v) + \frac{1}{K-C} \sum_{v \notin C} \hat{r}_{cp}^v \label{CP loss} \\
\mathcal{L}_{ncp} = & \frac{1}{S}\sum_{v \in S} (r_{ncp}^v - \hat{r}_{ncp}^v) + \frac{1}{K-S} \sum_{v \notin S} \hat{r}_{ncp}^v \label{NCP loss}
\end{align}

Here, mean absolute errors are computed separately for regions containing the plaque type and regions which do not contain the plaque type, thus weighing both regions equally.

\subsection{CAD-RADS Classification}
A quantitative description of cross-sectional area of the coronary artery is inferred directly by triangulating the $N_{\theta}$ nodes with the coronary artery centerline per MPR slice. This results in area measures up until lumen ($A_l$), NCP ($A_{l+ncp}$), and the outer wall ($A_{l+ncp+cp}$). These can be decomposed into $A_l$, $A_{cp}$, and $A_{ncp}$ through a simple subtraction step, resulting in area measures for lumen, CP, and NCP respectively. For a coronary artery of length $L$, this produces a 1D-signal $S = (A^1, A^2, \dots, A^L)$ for each of the three area measures. Signals are padded to a length of 512, which then serve as the input to the CAD-RADS classification CNN. For each patient, the CNN receives signals for the RCA, LAD, and LCx. Out of the predicted CAD-RADS categories for these signals, the highest prediction is presented as the final CAD-RADS score.

\begin{figure*}[b!]
\centering
     \includegraphics[trim=10 90 0 100, clip, width=1.02\textwidth]{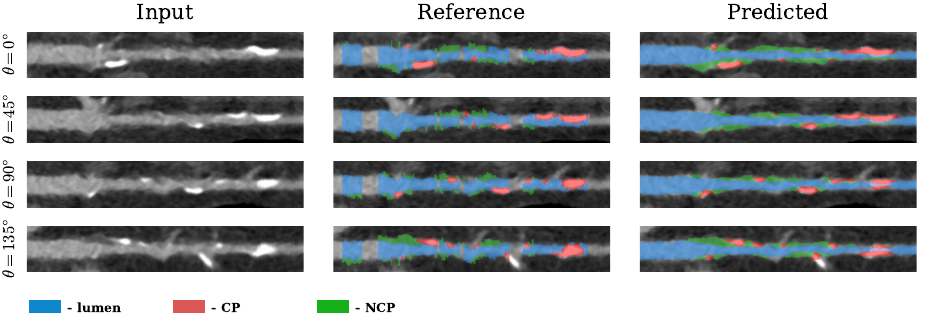}
     \caption{Example of plaque segmentation: The first column denotes a longitudinal view of} the MPR input volume, the second column shows the manually labeled reference segmentation, and the third column shows the automatically generated lumen and plaque segmentations obtained from the FanCNN. Note the uneven reference annotations resulting from the manual segmentation of cross-sectional artery slices. Results are presented at different angles $\theta$ around the coronary artery centerline. Also note how the coronary artery centerline tracing through the most distal CP causes a small gap in the automatic segmentation.
       \label{postprocess}
\end{figure*}

We choose to employ a small 1D ResNet architecture, of which the ability to capture local features along vessel centerlines is well-suited for CAD-RADS scoring. The network first features a convolution layer of kernel size 7 and stride 2, followed by a max-pooling layer of kernel size 3 and stride 2. This is followed by 3 Basic ResNet blocks, each containing a single convolutional layer pair. Each block performs two convolutions with kernel size 3, of which the first convolution is of stride 2. A max-pooling layer subsequently selects the maximum activation along the $L$ dimension for each channel, for a total of 64 features. This is then used as the input to a final fully-connected layer with 5 outputs and a sigmoid activation function, corresponding to the 5 CAD-RADS categories. All convolutional layers are followed by a batch normalization layer and a ReLU activation function. Furthermore, the first convolution produces 8 output channels, while the 3 ResNet blocks produce 16, 32, and 64 features respectively.

Similar to Denzinger \textit{et al.}, we incorporate the ordinal structure of CAD-RADS categories into the loss function directly by optimizing the network with the binary cross-entropy for each of the 5 outputs separately \cite{Denzinger2022}:

\begin{equation}
    \mathcal{L} = - \frac{1}{K}\sum_{j}^K\sum_{i}^5y_i^j\log (p_i^j) + (1-y_i^j)\log(1-p_i^j)
    \label{CADRADS loss}
\end{equation}

where $K$ is the batch size expressed in number of patients. This effectively transforms the classification task into a multi-label problem, where each CAD-RADS score is encoded by an $N$-hot vector of length 5, in which $N$ is the CAD-RADS score.

\section{Evaluation}
\label{EvalSec}
Automatic plaque segmentation was evaluated through plaque volume quantification on both lesion and artery level. For this, automatic plaque volumes were derived by voxelizing the automatically obtained meshes. Agreement between automatically and manually derived CP, NCP, and total plaque volumes were compared by assessing the intraclass correlation coefficient (ICC), lesion-level sensitivity, average number of true positive lesions (TP/art.) and false positive lesions (FP/art.), and a Bland-Altman analysis, for which the bias and limits of agreement were reported. Inter-observer agreement was evaluated using a Bland-Altman analysis. 

Quality of the automatic lumen segmentations was assessed by evaluating the method's performance on the Coronary Artery Stenoses Detection and Quantification Evaluation Framework \cite{Kirisli2013}. Performance was compared based on the Dice similarity coefficients (DSC), as well as the mean surface distance (MSD) and Hausdorff distance (HD) between automatic and reference lumen mehses.

The ability of our method to automatically assign patients to a correct CAD-RADS category was evaluated with the accuracy, the linearly weighted Cohen's kappa ($\kappa$), and the Matthew's correlation coefficient (MCC).

\section{Experiments and Results}
\label{Experiments}
\subsection{Coronary Artery Mesh Generation}
The 95 patients in the training set were used to train and validate automatic mesh generation in 5-fold cross-validation. All hyperparameter tuning was performed through preliminary experiments across these folds. Scans from the test set were not used in any way during development or parameter tuning.

During training, augmentation was performed to make the network more robust to the exact centerline location within the lumen. For this, centerline points were randomly shifted up to 0.6 mm within each cross-sectional plane before intensity profiles were extracted. Furthermore, input volumes were randomly flipped in both $N_{\theta}$ and $L$ direction with $p=0.5$. Rotation-based augmentation was excluded, considering that the data is implicitly invariant to such alterations.

To compensate for the unbalanced and sparse distribution of plaque in the datasets, the model was trained on batches with samples of four categories: no plaque, CP, NCP, and mixed plaque (see Section~\ref{meshSec}). While iterating over the training set, a subsection of an MPR volume was selected featuring one of these four categories to prevent the model from creating a bias towards the most common class in the dataset. MPR volume samples were of length 10 mm, with the category of interest being present in at least the central slice for each data sample (i.e. $z=5$ mm). Before selection of training patches, the input volume is padded with circular padding in the $N_{\theta}$ direction and mirror padding in the $L$ direction in order to preserve the original size of the full cylindrical MPR. The same padding is applied at test time, during which the FanCNN is able to process the full length of a coronary artery in a single forward pass due to its fully convolutional nature.

During training, a batch size of 32 was used to minimize the loss function. The network contained a total of 170K parameters and was updated for every batch with the AdamW optimizer \cite{Loshchilov2017}, employing an initial learning rate of 0.01. The network was trained for a total of 6,000 epochs, with the learning rate being multiplied with a factor of $\gamma = 0.1$ at epochs [3,000; 4,000; 5,000]. After cross-validation, the FanCNN was trained with all 95 patients for 5 different seeds resulting in an ensemble of 5 networks.

\begin{table*}[t]
\centering
\caption{Lesion-level quantification scores for different model versions. Results are listed per network type (NN) and by use of classifier attention (CA). Sensitivity, average number of true positive lesions (TP/art.) and false positive lesions per artery (FP/art.), and ICC between predicted and reference lesion volume are reported for CP, NCP, and total plaque.}
\label{Ablation}
\begin{tabular}{llllllllllllllllll}
\toprule
NN     & CA          & \multicolumn{4}{c}{CP}               & & \multicolumn{4}{c}{NCP}       & & \multicolumn{4}{c}{Total}  \\\cmidrule{3-6}\cmidrule{8-11}\cmidrule{13-16}
       &             & Sens.          & TP/art.      & FP/art.            & ICC             & & Sens.          & TP/art.      & FP/art.         & ICC             & & Sens.          & TP/art.      & FP/art.            & ICC \\
\midrule
FanCNN & \checkmark  & \textbf{92}\%  & \textbf{4.4} & 0.2                & \textbf{0.98}   & & 58\%           & 1.4          & 1.1             & \textbf{0.79}   & & 83\%           & 4.4          & \textbf{0.3}       & \textbf{0.85}\\          
FanCNN &             & 90\%           & 4.3          & 0.2                & 0.97            & & \textbf{83}\%  & \textbf{2.0} & 2.3             & 0.73            & & \textbf{87}\%  & \textbf{4.6} & \textbf{0.3}       & 0.77         \\      

GCN    & \checkmark  & 85\%           & 4.1          & 0.3                & 0.96            & & 42\%           & 1.0          & \textbf{0.8}    & 0.68            & & 75\%           & 4.0          & 0.4                & 0.62         \\
GCN    &             & \textbf{92}\%  & \textbf{4.4} & \textbf{0.1}       & \textbf{0.98}   & & 79\%           & 1.9          & 3.5             & 0.54            & & 85\%           & 4.5          & 1.0                & 0.54         \\  
\bottomrule
\end{tabular}
\end{table*}

\begin{figure*}[b]
     \centering
     \begin{subfigure}[b]{\textwidth}
         \centering
         \includegraphics[width=0.98\textwidth]{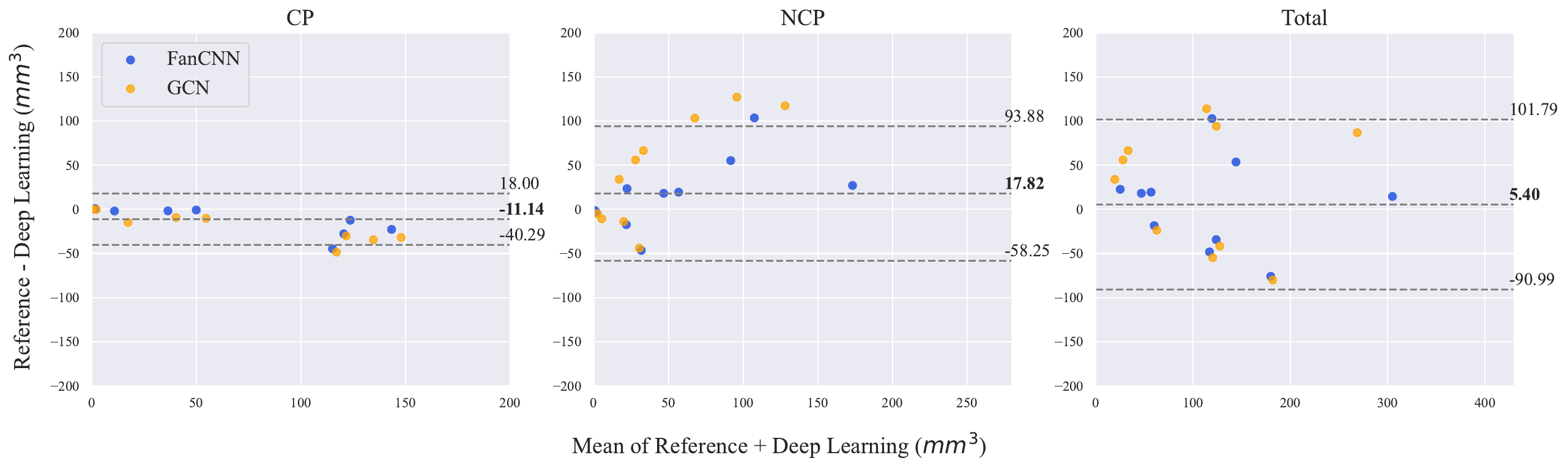}
     \end{subfigure}
     \hfill
     \begin{subfigure}[b]{\textwidth}
         \centering
         \includegraphics[width=0.98\textwidth]{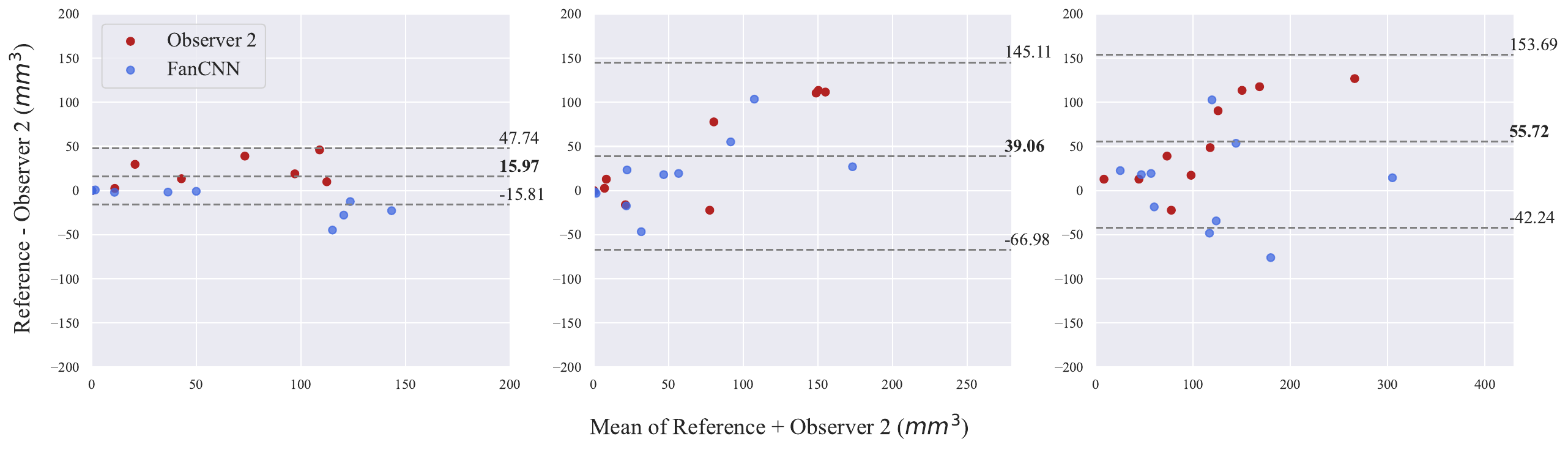}
     \end{subfigure}
     \caption{Vessel-level Bland-Altman analysis between reference and the automatic method (top) and between reference and Observer 2 (bottom). Bias and limits of agreement are presented for the FanCNN in the top row, and for inter-observer agreement in the bottom row.}
\label{blandaltman}
\end{figure*}

\subsection{Plaque Burden Quantification}
Automatically extracted lesion- and artery-level measurements were compared with the reference (Observer 1) annotations. Lesion-level results for coronary artery plaque quantification are presented in Table~\ref{Ablation}. To demonstrate the benefit of classifier attention, we present results with and without this inference step. In general, the networks perform similarly for CP and total plaque. The use of classifier attention led to fewer false positives. Furthermore, sensitivity levels remain similar for CP and total plaque, though a clear drop in sensitivity is observed for NCP. Classifier attention is of little impact on CP quantification, but improves the ICC for both NCP and total plaque quantification, indicating that only small lesions are typically missed and lesion sizes correlate well. An example of plaque segmentation is presented in Fig.~\ref{postprocess}.

\begin{table}[h]
\centering
\caption{Confusion matrix for patient-level stenosis categorization according to CAD-RADS (0: 0\%, 1: 1-24\%, 2: 25-49\%, 3: 50-69\%, 4: 70-99\%, 5: 100\%) in the hold-out and external test set, expert reference vs. automatic method. Linearly weighted $\kappa$ scores are reported for both sets.}
\label{Confusion matrix}
\begin{minipage}{\linewidth}
\centering
\begin{tabular}{c | rrrrrr | r}
\multicolumn{1}{c}{} & \multicolumn{6}{c}{\textit{Automatic}} \\
\textit{Reference $\downarrow$} & 0   & 1  & 2  & 3  & 4 & 5 & Total \\ \cmidrule{1-7}
0        & \textbf{121} & 17 & 0  & 0  & 0 & 0 & 138\\
1        & 11  & \textbf{40} & 10 & 1  & 0 & 0 &  62\\
2        & 3   & 14 & \textbf{32} & 7  & 6 & 0 &  62\\
3        & 0   & 1  & 6  & \textbf{14} & 2 & 0 &  23\\
4        & 0   & 0  & 1  & 4  & \textbf{7} & 0 &  12\\
5        & 0   & 0  & 0  & 0  & 3 & \textbf{0} &   3\\
\midrule
\multicolumn{1}{c}{Total} & 135 & 72 & 49 & 26 & 18 & 0 & 300 \\
\end{tabular}\par
\bigskip
(a) Hold-out test set ($\kappa$ = 0.75)
\end{minipage}

\bigskip

\begin{minipage}{\linewidth}
\centering
\begin{tabular}{c | rrrrrr | r}
\multicolumn{1}{c}{} & \multicolumn{6}{c}{\textit{Automatic}} \\
\textit{Reference $\downarrow$} & 0   & 1  & 2  & 3  & 4 & 5 & Total \\ \cmidrule{1-7}
0        & \textbf{259} & 54 &  1 & 0  & 0  & 0 & 314\\
1        & 11  & \textbf{76} & 55 & 16 & 0  & 0 & 158 \\
2        & 0   & 8  & \textbf{53} & 35 & 3  & 0 & 99\\
3        & 0   & 1  & 16 & \textbf{29} & 11 & 0 & 57\\
4        & 0   & 0  & 1  & 13 & \textbf{13} & 0 & 27\\
5        & 0   & 0  & 0  &  0 & 3 &  \textbf{0} & 3\\
\midrule
\multicolumn{1}{c}{Total} & 270 & 139 & 126 & 93 & 30 & 0 & 658 \\
\end{tabular}\par
\bigskip
(b) External test set ($\kappa$ = 0.71)
\end{minipage}

\end{table}

Fig.~\ref{blandaltman} shows artery-level Bland–Altman plots for the agreement between reference and automatically obtained plaque volumes, and between the observers. Automatically obtained volume tends to overestimate CP and underestimate NCP. For all plaque types, the second observer on average underestimated volume compared to the reference. Bland-Altman bias is closer to zero for the automatic method than for the second observer for all plaque types. For CP and total plaque, limits of agreement are similar for both the automatic method and the second observer. For NCP, limits of agreement are more narrow for the automatic method.

\subsection{CAD-RADS Optimization}
The CAD-RADS categorization was optimized using the 1,110 training scans from Set 1 in 5-fold cross-validation.

Since CAD-RADS only applies to vessels with a diameter larger than 1.5 mm, only coronary artery mesh predictions that met this criterion were used for training and testing \cite{Cury2016a}. Furthermore, lumen area measures along the coronary artery centerline were normalized to a percentage difference in area from one cross-section to the next to accentuate relative differences. CP and NCP areas were combined into one signal, representing total plaque along the coronary artery. Since labels were only available at the patient level, the method was trained by providing the model with the three main coronary artery signals (LAD, LCx, RCA). Only the highest output score was backpropagated, given that the most stenotic artery determines the final CAD-RADS score.

To train the CNN, parameter updates were performed with the AdamW optimizer \cite{Loshchilov2017} using an initial learning rate of 0.01 and a batch size of 256. The network was optimized for 12,000 epochs, during which the learning rate was updated by a factor of $\gamma = 0.1$ at epochs [6,000; 8,000; 10,000]. At test time, all arteries derived from the automatically extracted coronary artery tree were provided as an input to the CNN, and were processed by an ensemble of 5 differently seeded networks optimized on the full training set. All aforementioned methods and optimizations were implemented with the PyTorch deep learning library \cite{Paszke2019}.

\subsection{CAD-RADS Categorization}
The confusion matrices in Table~\ref{Confusion matrix} display the performance of the method on CAD-RADS categorization. Accuracy on the hold-out test set was 71\%, while one-class-off accuracy was 96\%. The linearly weighted $\kappa$ was 0.75. The external test set accuracy was 65\% while the one-class-off accuracy was 97\%, and the linear $\kappa$ was 0.71. The method tended to overestimate CAD-RADS severity more than underestimate it. CAD-RADS overestimation by more than one category was due to the coronary artery centerline passing through a calcification. Underestimation was due to a variety of reasons, which include the method underestimating an NCP lesion and label noise.

\section{Comparison with Other Methods}
\label{CompareSec}
\subsection{Plaque Burden Quantification}
\subsubsection{GCN}
We compared our plaque quantification method using a CNN with a GCN, which is commonly used to process mesh-oriented data \cite{Kipf2016}. As shown by Wolterink \textit{et al.} \cite{Wolterink2019a}, GCNs are an excellent choice to create a smooth tubular surface mesh for the coronary artery lumen. We extended this model to generate a plaque surface mesh for our purposes. In this format, intensity profiles are considered to be nodes that aggregate information from neighboring nodes through a graph convolution. To compare performance of the FanCNN with a GCN, the FanCNN architecture was translated to a GCN by replacing all 3D convolutions with graph attention (GATv2) layers \cite{Brody2021}. The rest of the model architecture remained unchanged, and the number of parameters for the GCN was kept roughly equal to that of the FanCNN. Table~\ref{Ablation} lists lesion detection results for the GCN-based approach. Though the reported results show excellent agreement for CP, the GCN fails to accurately identify and quantify NCP. NCP lesion volumes correlate poorly for both settings evaluated for the FanCNN and sensitivity is only high when the FP rate is high. All evaluated GCN settings were either outperformed by the corresponding FanCNN surrogate in terms of NCP and total plaque, or performed similarly in the case of CP. This is likely due to limited spatial awareness between neighboring vertices in GCNs, for which the aggregation function does not take directionality into account. Sense of direction appears to be of high importance for the identification of NCP, where a local combination of low-intensity values does not necessarily indicate NCP. This in contrast to CP, which can typically be identified directly by its distinctive attenuation range.

\begin{table}[t]
\centering
\caption{Comparison with previously published results on lesion-level plaque quantification. For each study, the number of scans in the test set (N) is reported, intra-class correlation coefficient (ICC), and Bland-Altman bias and limits of agreement in $mm^3$ (Bland-Altman). Results are specified per plaque type, and an asterisk denotes patient-level results.}
\label{Previous lesionwise}
\begin{tabular}{llrlr}
\toprule
 & Study & N & ICC  & Bland-Altman\\
\midrule
CP & \href{https://www.sciencedirect.com/science/article/pii/S1361841516300226?ref=cra_js_challenge&fr=RR-1}{Wolterink et al.}* \cite{Wolterink2016}  &  100     & 0.77 & -19.6 (-144.7, 105.4)   \\
   & \href{https://www.sciencedirect.com/science/article/pii/S0010482519302860}{Kigka et al.} \cite{kigka2019three}                                   &   18     & 0.93 &  -2.0 (  -5.7,   1.7)   \\
   & \href{https://www.sciencedirect.com/science/article/pii/S258975002200022X}{Lin et al.} \cite{Lin2022}                                            &  275     & 0.95 &   2.4 ( -43.6,  48.5)   \\
   & Ours                                                                                                                                             &   10     & 0.98 &  -2.6 (  -9.0,   3.8)   \\
\midrule
NCP    & \href{https://www.sciencedirect.com/science/article/pii/S0010482519302860}{Kigka et al.} \cite{kigka2019three}                               &   18     & 0.92 & -8.7 ( -87.8,  70.5)   \\
       & \href{https://www.sciencedirect.com/science/article/pii/S258975002200022X}{Lin et al.} \cite{Lin2022}                                        &  275     & 0.94 &  3.0 (-114.5, 120.4)   \\
       & Ours                                                                                                                              &   10     & 0.79 &  6.0 (-31.1,   43.1) \\
\midrule
Total & \href{https://www.sciencedirect.com/science/article/pii/S258975002200022X}{Lin et al.} \cite{Lin2022}                                   &  275     & 0.96 &  5.4 (-114.7, 125.6) \\
      & Ours                                                                                                                            &   10     & 0.85 & -0.8 (-30.7,   29.1) \\
\bottomrule
\end{tabular}
\end{table}

\begin{table}[h]
\centering
\caption{Quantitative scores for the training set of the Coronary Artery Stenoses Detection and Quantification Evaluation Framework. Dice similarity coefficient (DSC), mean surface distance (MSD), and Housdorff distance (HD) metrics are reported. A comparison is provided by listing several previously described methods with published results on the same set. Results are reported separately for healthy (H) and diseased (D) vessel segments.}
\label{lumen-seg}
\begin{tabular}{lcccccccc}
\toprule
 Study & \multicolumn{2}{c}{DSC} & &\multicolumn{2}{c}{MSD (mm)} & &\multicolumn{2}{c}{HD (mm)} \\ \cmidrule{2-3} \cmidrule{5-6} \cmidrule{8-9}
 & H & D && H & D && H & D \\ \midrule
\href{https://link.springer.com/chapter/10.1007/978-3-319-13972-2_13}{Lugauer et al.} \cite{lugauer2014precise}   & \textbf{0.77} & \textbf{0.75} & & 0.32 & \textbf{0.27} & & 2.79 & 1.96 \\
\href{https://aapm.onlinelibrary.wiley.com/doi/full/10.1002/mp.12121}{Freiman et al.} \cite{freiman2017improving}   & 0.69 & 0.74 & & 0.49 & 0.28 & & 1.69 & \textbf{1.22} \\
Mohr et al. \cite{Kirisli2013}      & 0.75 & 0.73 & & 0.45 & 0.29 & & 3.73 & 1.87 \\
\href{https://link.springer.com/chapter/10.1007/978-3-030-35817-4_8}{Wolterink et al.} \cite{Wolterink2019a} & 0.75 & 0.73 & & 0.25 & 0.28 & & 1.53 & 1.86 \\
Ours             & \textbf{0.77} & 0.72 & & \textbf{0.22} & 0.28 & & \textbf{1.29} & 1.54 \\ \bottomrule
\end{tabular}
\end{table}

\subsubsection{Previously published work}
We further compared the proposed plaque quantification with previously published methods (Table~\ref{Previous lesionwise}). For each method, we list, where available, the ICC between an observer and the proposed methodology, and the lesion-wise Bland-Altman bias and limits of agreement of the plaque type. Overall, Bland-Altman limits of agreement are narrower for the proposed method but have a lower ICC compared to Kigka \textit{et al.} \cite{kigka2019three} and Lin \textit{et al.} \cite{Lin2022}. Bland-Altman biases are similar for both methods. Given that previous works do not provide trained models and the methods are not publicly available, we compared performance as reported in the original publications. Note that training and test data used are from different sources and of various sizes. Hence, this comparison serves only as an indication.

\subsection{Lumen Segmentation}
A comparison of the proposed method's performance on lumen segmentation is provided in Table~\ref{lumen-seg}. We compare several state-of-the-art methods which all reported scores based on the same training set of 78 segments obtained from 18 CCTA images. The results indicate that our method achieves similar performance for all three metrics.

\subsection{CAD-RADS Categorization}
The proposed method was further compared with automatic CAD-RADS categorization published in previous work (Table~\ref{Previous work}). We compared, where available, the number of scans used to evaluate the study and the linearly weighted $\kappa$ score along with 95\% confidence interval (CI). Our method outperforms previous works on the six-class problem except for the work of Lin \textit{et al.}. However, the listed results cannot be directly compared, as different datasets of different sizes and distributions of CAD-RADS scores were used. 
The average reference CAD-RADS score in particular was significantly lower in our work compared to previous studies, which may impact the listed metrics.
The methods by Paul \textit{et al.} and Zreik \textit{et al.} directly infer the CAD-RADS score from MPR volumes without intermediate steps \cite{paul2022evaluation, Zreik2018}. Both methods by Denzinger \textit{et al.} aggregate features of MPR volumes from the full coronary artery tree to predict a patient-level CAD-RADS score \cite{Denzinger2020, Denzinger2022}.
Lin \textit{et al.} \cite{Lin2022} compute stenosis percentage directly in each artery based on lumen, CP, and NCP segmentations \cite{Lin2022}, and they do not evaluate CAD-RADS on patients diagnosed with no CAD (CAD-RADS 0).

\section{Discussion and Conclusion}
In this work, a method for automatic coronary artery lumen, CP, and NCP segmentation, plaque quantification, and subsequent CAD-RADS prediction in coronary CTA was presented. The method leverages the inherent tubular structure of coronary arteries to compute a surface mesh for coronary artery lumen, CP, and NCP boundaries by using a 3D CNN that estimates the radial extent of these structures. Unlike previous works on plaque quantification, the proposed method directly infers an anatomically plausible contiguous structure for lumen and plaque from a coronary artery centerline, and does not require any thresholding typically used in voxel-based segmentation. Output meshes inherently meet the requirements for downstream applications. Area measures of cross-sectional planes along the coronary artery centerline are calculated for the obtained meshes, which are processed by a 1D CNN to directly infer patient-level CAD-RADS category.

The results in Table~\ref{Ablation} reveal that lesion identification and quantification are excellent for CP and total plaque, and good for NCP. Vessel-level agreement of plaque quantification is shown to be on par with the second observer (Fig.~\ref{blandaltman}). Both results indicate the complexity of the NCP quantification, for which limits of agreement are much wider compared to CP quantification, while still performing similar to inter-observer variability. The high inter-observer variability further emphasizes the potential benefit of an independent reference standard as expert annotation might be of limited accuracy, and may require multiple longitudinal views of MPR volumes to compensate for the limited view during 2D axial annotation. By using automatically generated lumen and plaque meshes in CAD-RADS classification, we show that the method is able to identify the largest stenosis in the coronary arteries, as indicated by the agreement with the expert-defined reference.

A retrospective analysis of outliers in automatic CAD-RADS classification indicates that the identification of CAD-RADS 5 remains challenging, as well as the correct quantification of large NCPs. This is not surprising, as large plaques causing CAD-RADS 5 were underrepresented in training for both lumen and plaque segmentation and the CAD-RADS classification. Furthermore, the receptive field of the model was tuned such that it covered the largest NCP lesions in the training set, which may not have been large enough to accurately quantify plaque in patients with larger NCP lesions resulting in a high CAD-RADS category. The clinical impact of such error is not large, as further (invasive) assessment is performed for all patients assigned to categories higher than CAD-RADS 2. In future work, it may be interesting to enlarge the field of view of the network and use a more diverse training set with larger plaques.

\begin{figure}[b!]
\centering
     \includegraphics[width=0.49\textwidth]{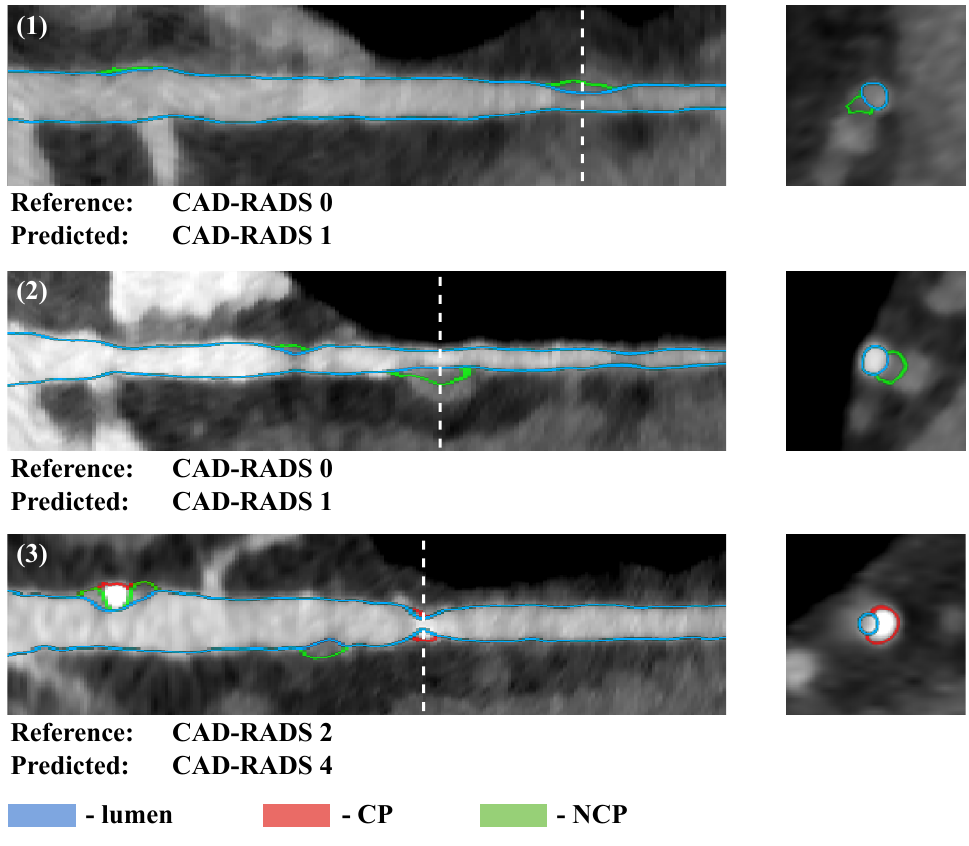}
      \caption{Examples of FanCNN failure cases, propagating to automatic CAD-RADS grading. For each example, a cross-section is provided at the failure site, indicated by a white dashed line in the longitudinal view of the MPR volume. (1) A bifurcation is wrongly identified as NCP, (2) a proximal vein causes an NCP-like CT density, and (3) the coronary artery centerline traces throug a CP, resulting in an overestimation of the stenosis severity.}
       \label{failure}
\end{figure}

A limitation of the current method is addressing the mixed plaque where NCP would be located distally to CP. Since the FanCNN builds the CP boundary on top of the NCP boundary, distal NCP would be missed. Though this scenario did not occur within the sets used, it is a plausible pathology and future research should address this. One may for example choose to re-frame the modeling approach to only generate a lumen and outer wall boundary, while identifying CP start- and end-points in a detached scheme. Calcium locations could subsequently be inserted into the predicted mesh, thus offering a more lenient inductive bias. Including such mixed plaque data in both training and testing would certainly be an interesting extension of the presented work. Furthermore, since no stent data was included in this study, this should be addressed in future work. A small amount of additional stent data as well as an additional classification head could address this issue, as the distinctive shape of stents should yield a well-defined optimization objective.

The current study's model architecture shares similarities with the approach presented in Alblas et al. \cite{alblas2022deep}. However, there are notable differences in terms of data, problem definition, and methodology. First, while Alblas et al. \cite{alblas2022deep} analyzed well-defined carotid arteries in black-blood MRI, our study focuses on coronary arteries in CCTA, where the coronary wall analysis is limited by CCTA resolution. Second, the regression of the outer wall thickness with a smooth L1 loss was sufficient for Alblas et al. \cite{alblas2022deep}, but the present study's focus on analyzing plaque components in the outer wall required a more complex approach. Since plaque is not present in every intensity profile, the study proposes the use of weighted L1 losses in combination with classifier attention. Additionally, while Alblas et al. \cite{alblas2022deep} noted a drop in algorithm performance in the presence of pathology, this study explicitly focuses on analyzing the presence of pathology in coronary arteries. Such differences highlight the requirement for tailor-made solutions to specific research questions and datasets.

\begin{table}[t]
\centering
\caption{Comparison of automatic CAD-RADS categorization methods. N describes the number of test set scans, $\kappa$ is the linearly weighted kappa score, and MCC describes the Matthew's correlation coefficient. Scores are presented for CAD-RADS 0 vs. 1-5, CAD-RADS 0-2 vs. 3-5, the three class case (0, 1-2, and 3-5), and for all six CAD-RADS grades separately.}

\label{Previous work}
\begin{tabular}{llrll}
\toprule
& Study & N & $\kappa$ (95\% CI) & MCC\\
\midrule
Two-class & \href{https://ieeexplore.ieee.org/abstract/document/8550784}{Zreik et al.} \cite{Zreik2018}                &  65  & 0.67 (0.45, 0.90) & 0.67\\
(0 vs. 1-5)         & \href{https://link.springer.com/chapter/10.1007/978-3-030-59725-2_5}{Denzinger et al.} \cite{Denzinger2020}& 955  & 0.52 (0.44, 0.60) & 0.52\\
         & \href{https://www.sciencedirect.com/science/article/pii/S2211568422000043?via%3Dihub}{Paul et al.} \cite{paul2022evaluation}  &  53  & 0.63 (0.43, 0.84) & 0.66\\
         & \href{https://proceedings.mlr.press/v172/denzinger22a.html}{Denzinger et al.} \cite{Denzinger2022}         & 976 & -                     & 0.62\\
         & Ours                                                                                                       & 300  & 0.79 (0.72, 0.86) & 0.79\\
\midrule
Two-class & \href{https://ieeexplore.ieee.org/abstract/document/8550784}{Zreik et al.} \cite{Zreik2018}              &  65   & 0.67\\
 (0-2 vs. 3-5)         & \href{https://link.springer.com/chapter/10.1007/978-3-030-59725-2_5}{Denzinger et al.} \cite{Denzinger2020}& 955  & 0.68 (0.64, 0.72) & 0.68\\
         & \href{https://www.sciencedirect.com/science/article/pii/S258975002200022X}{Lin et al.} \cite{Lin2022}      &  50  & 0.88 (0.75, 1)     & 0.89\\
         & \href{https://www.sciencedirect.com/science/article/pii/S2211568422000043?via%3Dihub}{Paul et al.} \cite{paul2022evaluation}  &  53  & 0.91 (0.78, 1)     & 0.91\\
         & \href{https://proceedings.mlr.press/v172/denzinger22a.html}{Denzinger et al.} \cite{Denzinger2022}         & 976  & -                    & 0.73\\
         & Ours                                                                                                       & 300  & 0.69 (0.57, 0.81) & 0.69\\
\midrule
Three-class & \href{https://ieeexplore.ieee.org/abstract/document/8550784}{Zreik et al.} \cite{Zreik2018}                &  65  & 0.67 (0.52, 0.82) & 0.62\\
            & \href{https://link.springer.com/chapter/10.1007/978-3-030-59725-2_5}{Denzinger et al.} \cite{Denzinger2020}& 955  & 0.63 (0.59, 0.67) & 0.58\\
            & \href{https://www.sciencedirect.com/science/article/pii/S2211568422000043?via%3Dihub}{Paul et al.} \cite{paul2022evaluation}  &  53  & 0.76 (0.63, 0.89) & 0.71\\
            & Ours                                                                                                       & 300  & 0.76 (0.70, 0.82) & 0.71\\
\midrule
Six-class   & \href{https://link.springer.com/chapter/10.1007/978-3-030-59725-2_5}{Denzinger et al.} \cite{Denzinger2020}& 955 & 0.62 (0.59, 0.65) & 0.42\\
            & \href{https://www.sciencedirect.com/science/article/pii/S258975002200022X}{Lin et al.} \cite{Lin2022}      &  50 & 0.87 (0.77, 0.96) & 0.80\\
            & \href{https://proceedings.mlr.press/v172/denzinger22a.html}{Denzinger et al.} \cite{Denzinger2022}         & 976 & -                    & 0.49\\
            & Ours                                                                                                       & 300 & 0.75 (0.70, 0.79) & 0.59\\
            & Ours (external)                                                                                            & 658 & 0.71 (0.68, 0.74) & 0.52\\
\bottomrule
\end{tabular}
\label{tab}
\end{table}

Since our proposed method generates meshes at the artery level, both bifurcations and veins in close proximity to the central coronary artery can prove detrimental to the detection and quantification of NCP, as image intensity ranges appear similar to coronary wall and plaque tissue (Fig. \ref{failure}). Furthermore, NCP lesions are not mutually exclusive with bifurcations, emphasizing the need for a model capable of interpreting a larger field of view in order to correctly identify plaque. In this work, we account for this using classifier attention, which reduces FPs by half while improving the ICC between reference and automatically quantified NCP volume (Table~\ref{Ablation}).

Automatic CAD-RADS scores which were off by more than one category were typically due to the coronary artery centerline temporarily diverging from the coronary artery or tracing through a plaque in the outer wall. Since the method operates directly on the centerline, any plaque quantification errors created due to inaccurate centerline extraction will propagate to inaccurate CAD-RADS scoring. One of the strongest assumptions this network makes is that the centerline has tracked through the coronary artery lumen, allowing it to build a hierarchical mesh of lumen and plaque. As such, mesh generation fails if this assumption does not hold. Tracing through a CP, which is typically caused by blooming artifacts, results in an overestimation of the CAD-RADS score (Fig. \ref{failure}), as the algorithm is likely to only detect calcifications without any identifiable lumen. Large NCP lesions on the other hand were underestimated, as the algorithm tends to identify the neighborhood as low-intensity lumen, thus not detecting the plaque. Furthermore, CAD-RADS 5 is particularly difficult to identify as the centerline extraction algorithm is unable to trace beyond total occlusions, resulting in an underestimation of stenosis severity. It is possible to correct such centerline errors with for example the manual correction performed by Lin \textit{et al.} \cite{Lin2022}, which will likely lead to improvement. Accounting for such large misclassifications in CAD-RADS would enhance the reliability and usage of the algorithm in a clinical scenario. Automatic centerline refinement strategies may further be investigated, such as correcting points with outlier HU values. Though we further include failure of the centerline extraction algorithm as an exclusion criterion, this only occurred in 2/2407 total scans. In both cases centerlines were identified, but were automatically discarded due to the presence of an abnormal ostium, which the method did not recognize. At test time one may consider an interactive scenario where, prior to the artery analysis, optional manual seed points may be placed for automatic centerline extraction \cite{Wolterink2019}.

Though the CAD-RADS categorization network architecture was not extensively evaluated in this work, some other architectures were considered. We experimented with a 1D Vision Transformer \cite{beyer2022better}, for which the results proved to be substantially worse compared to the 1D ResNet architecture (six-class hold-out $\kappa=0.63$). Our intuition is that Transformers are designed to capture global dependencies between elements in a sequence at an early stage, which may not be necessary for this particular task. The largest stenosis in an artery is typically determined by local factors, such as plaque buildup or narrowing, rather than global characteristics of the entire vessel. Future work may consider more extensive experimentation for the longitudinal accumulation of area measures along the coronary artery.

The method has been evaluated on a representative set of 958 scans with suspected CAD made on three different scanner types from two vendors. This study includes patients from both the Netherlands and China. The results show a high $\kappa$ for both the hold-out and external test set, with only 3.5\% of all scans being classified with a CAD-RADS category that differed from the reference CAD-RADS by more than one. Hence, the method is robust to these differences in data properties and acquisition. Despite this, it should be noted that specific combinations of reconstruction parameters may impact algorithm performance, typically as a result of high noise levels or stack artifacts \cite{denzinger2023scan}. Furthermore, CCTA volumes with limited resolution may result in partial volume effects, which unfortunately limit the quality of plaque quantification, especially in the case of small and mixed plaques. In future work, it may prove interesting to evaluate the method on a large dataset with a uniform distribution of CAD-RADS scores. Training the algorithm on a larger set of patients with higher CAD-RADS categories may further yield a higher accuracy. Such examples were heavily underrepresented in our data, with CAD-RADS 3-5 comprising only $13\%$ of all data.

Accurate segmentation of coronary artery plaque allows further analysis of plaque characteristics, such as identification of high-risk plaque or the quantification of mixed plaque, an important and independent predictor of future cardiac events. Given the complexity and low reproducibility of manual voxel-wise annotation of mixed plaque, we did not evaluate mixed plaque. Nevertheless, the method detects it as a prediction of both plaque types along the same ray. In future work, it may be interesting to explicitly evaluate the performance on mixed plaque quantification.

To conclude, this study presented an algorithm for the segmentation of coronary artery lumen and plaque and subsequent automatic CAD-RADS categorization by generating a mesh with a CNN operating on cylindrical data. To the best of our knowledge, this is the first method to automatically and accurately distinguish lumen, CP, and NCP with the guarantee of anatomically plausible shapes. The method allows automation of the complete clinical workflow from plaque evaluation to stenosis grading.

\label{DiscSec}
\bibliographystyle{IEEEtran}
\bibliography{tmi}

\end{document}